\documentclass[twocolumn,showpacs,preprintnumbers,amsmath,amssymb]{revtex4}

\usepackage{graphicx}
\usepackage{dcolumn}
\usepackage{bm}

\begin{document}


\title{Electronic spin transport in graphene field effect transistors}

\author{M. Popinciuc$^1$\footnote{Present address: II. Physikalisches Institut A, RWTH Aachen University, Germany} \footnote{Electronic address: popinciuc@physik.rwth-aachen.de}, C. J\'{o}zsa$^2$, P. J. Zomer$^2$, N. Tombros$^2$, A. Veligura$^2$, H. T. Jonkman$^1$, and B. J. van Wees$^2$}

\affiliation{$^1$Molecular Electronics and $^2$Physics of Nanodevices, Zernike Institute for Advanced Materials, University of Groningen, The Netherlands}

\date{\today}

\begin{abstract}
Spin transport experiments in graphene, a single layer of carbon atoms, indicate spin relaxation times that are significantly shorter than the theoretical predictions. We investigate experimentally whether these short spin relaxation times are due to extrinsic factors, such as spin relaxation caused by low impedance contacts, enhanced spin flip processes at the device edges or the presence of an aluminium oxide layer on top of graphene in some samples. Lateral spin valve devices using a field effect transistor geometry allowed for the investigation of the spin relaxation as a function of the charge density, going continuously from metallic hole to electron conduction (charge densities of $n\sim 10^{12}$cm$^{-2}$) via the Dirac charge neutrality point ($n\sim0$). The results are quantitatively described by a one dimensional spin diffusion model where the spin relaxation via the contacts is taken into account. Spin valve experiments for various injector/detector separations and spin precession experiments reveal that the longitudinal (T$_1$) and the transversal (T$_2$) relaxation times are similar. The anisotropy of the spin relaxation times $\tau_\parallel$ and $\tau_\perp$, when the spins are injected parallel or perpendicular to the graphene plane, indicates that the effective spin orbit fields do not lie exclusively in the two dimensional graphene plane. Furthermore, the proportionality between the spin relaxation time and the momentum relaxation time indicates that the spin relaxation mechanism is of the Elliott-Yafet type. For carrier mobilities of 2-5$\times 10^3$~cm$^2$/Vs and for graphene flakes of 0.1-2~$\mu$m in width, we found spin relaxation times of the order of 50-200~ps, times which appear not to be determined by the extrinsic factors mentioned above.

\end{abstract}

\maketitle

\section{Introduction}

Spintronics is the field of electronics that uses the spin of the charge carrier for the transport and manipulation of information. Preserving a spin state is fundamentally limited by the spin orbit interaction which provides the mechanism(s) for spin relaxation. Carbon based material systems promise a weak spin orbit interaction (i.e., long spin relaxation times) due to the low atomic number of carbon. Spin transport investigations have been reported for molecular systems and carbon nanotubes, see \cite{Naber07} for a review, and more recently for single or multi-layer graphene \cite{Hill06,Tombros07,Cho07, Nishioka07,Ohishi07,Wang08,Tombros08,Jozsa08,Goto08,Shiraishi08,Jozsa09,Han09a,Han09b}. Combined with the relatively large carrier mobilities \cite{Novoselov04,RiseofG,Bolotin08}, this determines a long spin relaxation length, i.e. a long distance which a spin can travel without loosing its initial orientation.

In previous experiments we deduced spin relaxation times of 150~ps (spin relaxation lengths of 1.5-2 $\mu$m) at room temperature \cite{Tombros07}, value confirmed by more recent findings \cite{Shiraishi08,Han09b}. These relaxation times are shorter than expected by at least one order of magnitude, pointing to an extrinsic spin relaxation mechanism rather than to an intrinsic one. In this study, we present all electrical spin valve and spin precession experiments aiming at determining the possible causes for such short spin relaxation times. From the experimental point of view, three major aspects are identified as discussed below. For theoretical investigations of the possible spin relaxation mechanisms we refer to \cite{Kane05,Hernando06,Min06,Yao07,Hernando08,Gmitra09,Ertler09,CastroNeto09}.

First, in order to combat the conductivity mismatch problem \cite{Schmidt00} and realize efficient electrical spin injection and detection in graphene, a thin aluminium oxide layer has been introduced in between the ferromagnetic Co injector/detector and graphene \cite{Schmidt00,Rashba00,Fert01,Smith01,Takahashi03,Staa08}. Yet, in some experiments the contact resistances are comparable in magnitude with the graphene square resistance over one spin relaxation length and significant spin relaxation occurs via the contacts. We quantify this effect by using an one dimensional spin diffusion model. Second, in all our previous experiments the whole graphene layer was covered by the thin aluminum oxide layer, a procedure which simplified the fabrication process. Whether this oxide layer, which needs to be only underneath the Co electrodes, induces extra spin relaxation was not known. In order to investigate this issue we performed experiments where the aluminium oxide barriers were present only underneath the Co electrodes. The third aspect is whether the low spin relaxation times are due to spin scattering at the edges of the graphene flakes. Therefore, we performed spin transport experiments using narrow structures (down to 100~nm in width, 15-20 times narrower than the spin relaxation length).

This paper is organized as follows. In Section~\ref{sec:Exp}, after a short introduction regarding the functioning of a four terminal spin valve device, we present fabrication/measurement details and general electrical characteristics of the devices. In Section~\ref{sec:ThExp} we discuss modeling and experiments pertaining the spin valve and spin precession experiments. Details of the modeling of spin relaxation via the contacts are presented in the Appendix.

\section{Experiments}\label{sec:Exp}
\subsection{Graphene lateral spin valve devices}\label{ssec:Basics}

\begin{figure}[h!]
  \includegraphics[width=\columnwidth]{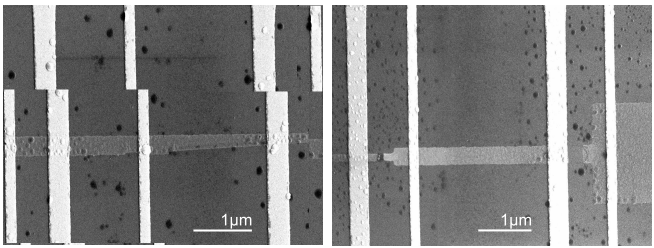}\\
  \caption{SEM of two graphene spin valves (type~III devices, fabricated by etching).} \label{fig:ExpSEM}
\end{figure}

In Fig.~\ref{fig:ExpSEM} we show scanning electron micrographs (SEM) of two graphene flakes contacted by four ferromagnetic Co electrodes (25-40~nm in thickness and different widths of up to 800~nm). All devices were fabricated on an oxidized silicon substrate and have an aluminium oxide layer all over the graphene (type~I devices) or only underneath the Co electrodes (type~II and type~III devices). The fabrication procedure which defines the device types is described in the next section.

\begin{figure}[h!]
  \includegraphics[width=0.8\columnwidth]{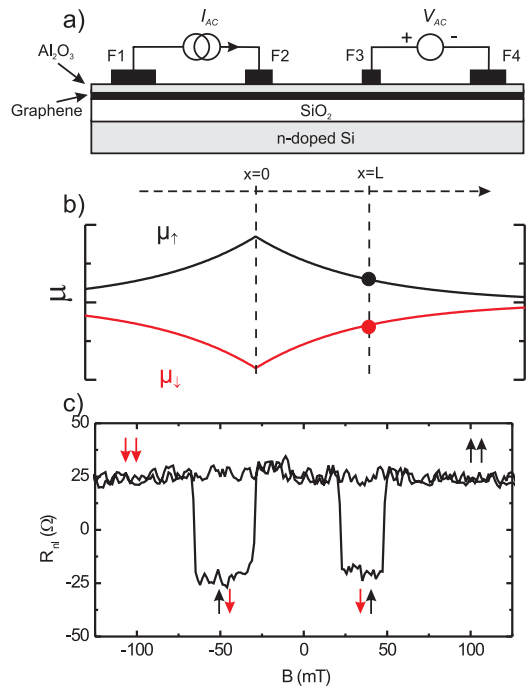}\\
  \caption{a) Schematic cross-section of a type~I graphene spin valve device measured in the non-local geometry. b) Representation of the electrical spin injection/detection process. c) Spin valve measurement (device I.A) showing a 50~$\Omega$ spin signal. Note the sharp switching and the symmetry of the spin signal around zero non-local resistance.}\label{fig:SVClassic}
\end{figure}

In Fig.~\ref{fig:SVClassic}a) we show a schematic cross section of a lateral graphene type~I spin valve device in the non-local measurement geometry \cite{Johnson88,Jedema02}.
For simplicity, we consider the outer electrodes (F1 and F4) as being non-magnetic. The current set between F2 and F1 creates a spin accumulation (imbalance) in graphene at the position of F2, i.e. the electrochemical potentials for the spin up and spin down ($\mu_\uparrow,\mu_\downarrow$) channels split \cite{vanSon87}, Fig.~\ref{fig:SVClassic}b). The spin accumulation $\mu_s=\mu_\uparrow- \mu_\downarrow$ diffuses away from the injection point ($x=0$) and, in the linear regime, obeys
the Bloch equation
\begin{equation}\label{eq:Diff}
D\nabla^2\mbox{\boldmath$\mu$}_s-\frac{\mbox{\boldmath$\mu$}_s}{\tau}+ \mbox{\boldmath$\omega$}\times\mbox{\boldmath$\mu$}_s=0,
\end{equation}
where $D$ and $\tau$ represent the spin diffusion constant and the spin relaxation time, respectively. The term $\mbox{\boldmath$\omega$}\times\mbox{\boldmath$\mu$}_s$ describes the precession of the spin accumulation in an external magnetic field $\mbox{\boldmath$B$}$ with the Larmor frequency $\mbox{\boldmath$\omega$}=g\mu_B\mbox{\boldmath$B$}$, with $g=2$ the gyromagnetic factor and $\mu_B$ the Bohr magneton. In Eq.~\ref{eq:Diff} we used a single relaxation time $\tau$ for reasons discussed later (see also \cite{Fabian07,Tombros08}).

The spin accumulation is probed non-locally by the voltage difference $V_{nl}$ between F3 and F4, with F3 placed at a distance $x=L$ of the order of $\lambda$ with respect to the injection point. Depending on its magnetization orientation (parallel/antiparallel) with respect to the spin accumulation, electrode F3 is sensitive to the electrochemical potential of either the spin up channel or the spin down one, Fig.~\ref{fig:SVClassic}b). The \textit{spin signal} is defined as the non-local resistance $R_{nl}=V_{nl}/I$, where $I$ represents the injection current. In the absence of precession $\mbox{\boldmath$\omega$}\times\mbox{\boldmath$\mu$}_s=0$ the spin accumulation decays exponentially with the distance with the characteristic length $\lambda =\sqrt{D\tau}$, the spin relaxation length. Under the assumption of high impedance ferromagnets one can show that
\begin{equation}\label{eq:Rnl}
R_{nl}=\pm \frac{P^2 R_{sq} \lambda}{2W}\exp{(-\frac{L}{\lambda})}
\end{equation}
where Eq.~1 from \cite{Jedema02} has been adapted for the two dimensional graphene. $R_{sq}$ represents the graphene square (sheet) resistance, $W$ is the width of the flake and $P$ is the spin injection/detection efficiencies of the ferromagnetic electrodes. The $+(-)$ signs correspond to the parallel (antiparallel) orientation of the magnetization of the injector (F2) and the detector (F3). The spin valve measurement, see Fig.~\ref{fig:SVClassic}c), consists in monitoring the non-local resistance as a function of an external magnetic field which manipulates the relative orientation of the injector/detector magnetizations. The electrodes are engineered to have different widths so that, due to the shape anisotropy, they switch their magnetization orientation at different magnetic fields applied along their easy axis (the length). The \textit{spin valve signal} $\Delta R_{nl}=2 |R_{nl}|$ represents the change in resistance when the injector/detector magnetization configuration changes from parallel to antiparallel. The exponential dependence of the spin valve signal $\Delta R_{nl}$ on the injector-detector separation $L$ allows for the extraction of the spin relaxation length, the parameter of most interest.

In the non-local geometry, due to the separation of the injection and detection circuits, magnetoresistance effects related to the charged current flow, such as anisotropic magnetoresistance and Hall effects, do not superimpose on the spin signal. Theoretically, the non-local resitance is determined only by the spin accumulation injected at $x=0$ which diffused towards the detector ($x=L$), i.e., there are no other resistances (voltages) on top of the non-local resistance (voltage). In most samples, however, a small background (frequency and gate voltage dependent) is measured. In a local measurement geometry, the injection and detection circuits share the same pair of electrodes. The spin signals are superimposed on the graphene resistance (typically few k$\Omega$) and are more difficult to detect, especially if the contact or graphene resistance fluctuates. Nevertheless, there are a few local measurements reported \cite{Tombros07,Nishioka07,Shiraishi08}.

\subsection{Device fabrication and electrical characteristics}\label{ssec:Fabrication}

The devices were fabricated on a dry oxidized, highly doped ($\rho$=1m$\Omega$cm) silicon substrate commercially available. The oxide was 500~nm or 300~nm thick. The application of a back gate voltage allows for the control of the charge carrier type and density in the graphene flakes. The gate electrode was fabricated by etching the back side oxide of the wafer followed by the deposition of a 100~nm thick Ti/Au layer. On this substrate we defined a set of Ti/Au markers using electron beam lithography (EBL). Next, the substrate was cleaned by oxygen plasma in order to remove the polymeric residuals from the EBL step. Then, highly oriented pyrolytic graphite (HOPG from GE Advanced Ceramics and SPI) was cleaved with a scotch tape and firmly pressed against the substrate. The tape is pealed off the substrate leaving a random distribution of graphite pieces among which there are also single layer graphene flakes. Using an optical microscope, the lowest contrast flakes were selected as potential single layer candidates. After that, the single layers were identified using atomic force microscope (AFM) measurements. At the same time their position with respect to the known position of the Ti/Au markers is noted which later on allowed for a precise positioning of the ferromagnetic electrodes. In a few cases Raman spectroscopy was performed in order to confirm the single layer selection performed by AFM \cite{Tombros08t}. The fabrication procedure continued differently for the different types of devices as explained below.

\begin{figure}[h!]
  \includegraphics[width=.75\columnwidth]{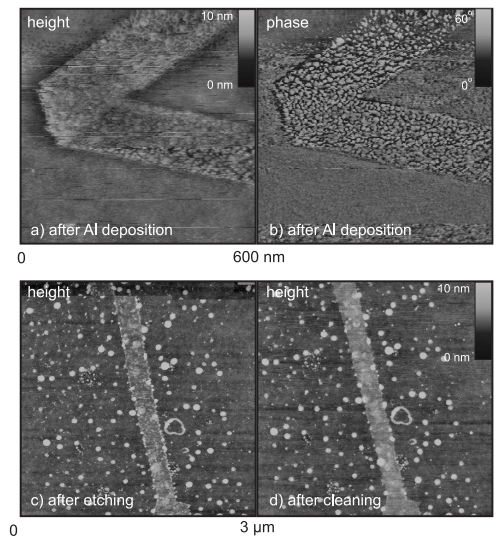}\\
  \caption{AFM measurements of two graphene flakes. a) and b) height and phase data ($600\times600$~nm$^2$) after Al deposition (type~I device). c) and d) height data ($3\times3~\mu$m$^2$) after etching and after annealing in Ar+H$_2$ atmosphere (type~III device).}
  \label{fig:ExpAFM}
\end{figure}

For the type~I samples, we continued with the thermal evaporation of a thin layer of Al (6~\AA) in an ultra high vacuum (UHV) system with the substrate being liquid nitrogen cooled. After that, the Al layer was oxidized for minimum 30 minutes in the load lock of the system in an oxygen atmosphere of about 100~mbar. Next, ferromagnetic electrodes of different widths were defined using EBL followed by electron beam evaporation of Co in a high vacuum system. After lift off in hot acetone for typically 10 minutes, the sample was glued on a chip carrier and electrical connections between the sample and the chip carrier were made using ultrasonic wire bonding. The sample was then placed in a vacuum container and measured. For the type~I devices, the graphene flakes are completely covered by the thin aluminium oxide. In Fig.~\ref{fig:ExpAFM}a)-b) we show AFM measurements of a graphene flake after the aluminium deposition. In spite of the cryogenic evaporation of Al, the oxide layer on graphene is granular in nature. SEM images, not shown here, support this observation.

In order to determine the influence of the AlOx layer covering graphene on the spin transport properties, devices where the AlOx layer was deposited only underneath the Co electrodes were fabricated (type~II devices). In this case an EBL step was performed before the sequential deposition/oxidation of Al and the thermal evaporation of Co in the same UHV system. Due to the configuration of the UHV system the Al deposition on the cooled stage was done under an angle of 30$^\circ$ with respect to the sample normal. The deposition was set to take place along the electrode length with a precision of about 1-2$^\circ$. This could potentially result in shadowing effects due to the resist walls, yielding non-uniform deposition of Al in the patterned areas.

The type~III devices were fabricated in order to investigate spin transport in narrow graphene ribbons. In this case all samples have the AlOx layer only underneath the Co contacts (same as type~II). However, naturally narrow flakes are difficult to spot under an optical microscope. To fabricate devices with widths down to 100~nm, somewhat larger flakes were selected and then oxygen plasma etched, prior to the definition of the composite AlOx/Co electrodes. The oxygen plasma etching step often resulted in a contaminated surface. Even though after etching the flakes were annealed in an Ar(95)\%:H$_2$(5\%) atmosphere at 350$^\circ$ for about two hours, large particles of unknown chemical composition still remain on the surface, see Fig.~\ref{fig:ExpAFM}c)-d).

The electrical measurements were performed using a standard AC lock-in technique ($f\leq$17~Hz) in the four terminal non-local geometry. The AC current (1-20~$\mu$A) was set between a pair of electrodes and the in phase AC non-local voltage was measured between the other pair of electrodes. The room temperature measurements were performed with the sample inserted in a vacuum container, base pressure of $\sim5\times 10^{-6}$~mbar (measured close to the pump), placed in between the poles of a room temperature electromagnet. The measurements were done in vacuum in order to reduce the hysteresis of the graphene resistance as a function of the gate voltage and to avoid the degradation of the spin signal as a function of time. In the case of atmospheric measurement conditions, the reduction of the spin signal in time (on a time scale of about 24 hours) was accompanied by an increase of the contact resistances. The low temperature (4.2~K) measurements were performed in an Oxford cryostat with the sample placed in a dip-stick and using a superconducting magnet.

\begin{figure}[t!]
  \includegraphics[width=\columnwidth]{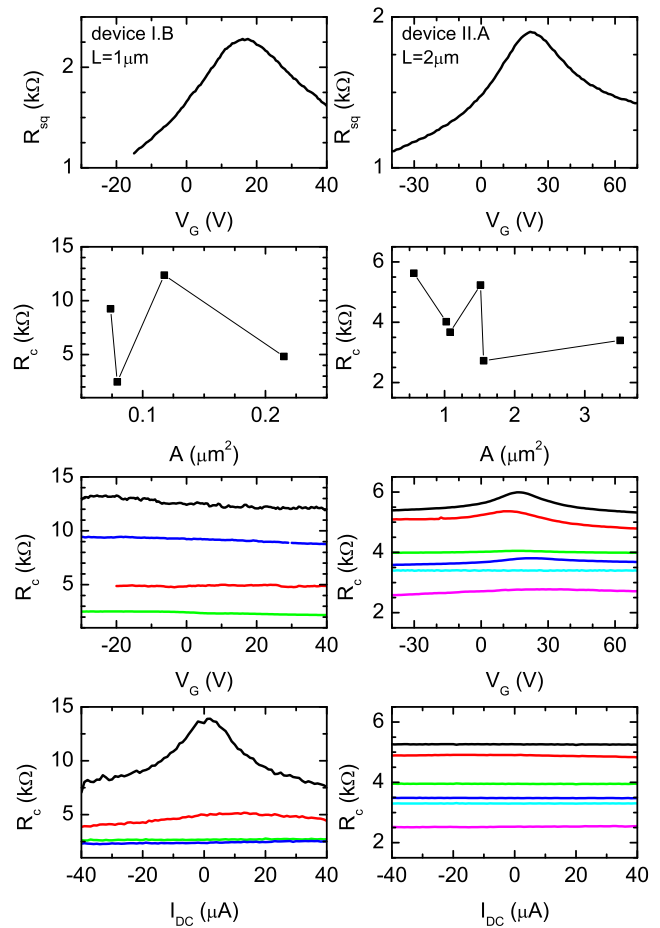}\\
  \caption{Graphene and contact electrical characteristics (differential resistance $dV/dI$) for type~I (left column) and type II (right column) devices.}
  \label{fig:ExpProp}
\end{figure}

In Fig.~\ref{fig:ExpProp} we compare electrical characteristics of graphene and contacts for two devices, one type~I and one type~II. As a function of the gate voltage, the four terminal measurements of the graphene resistivity (top panels) show a maximum which corresponds to the Dirac neutrality point where the average charge density is equal to zero. At negative voltages with respect to the Dirac point position the conduction takes place via holes, whereas for positive voltages the conduction happens via electrons. The mobility of all samples presented in this study was in the range 2-5$\times 10^3$~cm$^2$/Vs. A clear correlation between the carrier mobilities of the type~I (completely covered by aluminium oxide) and type~II samples (oxide barriers only underneath the electrodes) could not be established. For the type~III devices the carrier mobilities were the lowest (2-2.5$\times 10^3$~cm$^2$/Vs).

In the lower panels of Fig.~\ref{fig:ExpProp} we show three-terminal contact  resistance measurements as a function of contact area, gate voltage and DC current bias. In general, the contact resistances of type~I and type~III devices show a large spread in values and no clear scaling with the contact area. For the type~II samples, the spread is less and there seems to be some scaling of the contact resistance with the area. For all types of contacts, a small and irregular gate voltage dependence of the contact resistance is recorded. Some low resistance contacts show a similar behavior to graphene, i.e. the contact resistance is higher when the graphene flake is at the charge neutrality point. Most likely, this indicates that part of the graphene layer contributes to the contact resistance. With respect to the DC bias current dependence, some high resistance contacts show high values in zero bias conditions whereas low resistance contacts show very weak or no DC bias dependence. Unfortunately, the large spread in the overall behavior does not allow for a quantitative analysis. We take these observations and the AFM measurements (Fig.~\ref{fig:ExpAFM}) as evidence that the contacts characteristics are determined by a random distribution of regions with increased transparency (thinner oxide layer or pinholes) in contrast to the ideal case of tunnel barriers.

\section{Modeling and experiments}\label{sec:ThExp}
\subsection{Modeling the contact resistance effects in the spin valve measurements}\label{ssec:ThRc}

Analyzing Fig.~\ref{fig:SVClassic}a) we see that the spins injected by F2 do not only diffuse symmetrically to the left and to the right in graphene, but can also return into the ferromagnetic electrode where they loose their initial orientation very fast (due to the strong spin orbit coupling in the ferromagnet). Therefore, the magnitude of the spin accumulation created in graphene is drastically reduced if the contact resistance is much lower than the graphene resistance over one spin relaxation length. This effect is known as the conductivity mismatch problem \cite{Schmidt00}. Taking into account the spin current drawn by the electrode, see the Appendix for details, one can show that
\begin{equation}
R_{nl}=\pm\frac{2P^2 R_{sq}\lambda}{W} \frac{(R/\lambda)^2\exp(-L/\lambda)}
{(1+2R/\lambda)^2-\exp(-2L/\lambda)}, \label{eq:Rnl2}
\end{equation}
where the parameter $R$ given by
\begin{equation}
R=\frac{R_C}{R_{sq}}W
\end{equation}
represents spin relaxation due to the finite contact resistances $R_C$ of the injector (F2) and detector (F3). In fact, $R/\lambda$ represents the ratio between the contact resistance and the graphene resistance over one spin relaxation length. Eq.~\ref{eq:Rnl2} is in agreement with \cite{Takahashi03} given our assumptions of negligible ferromagnet resistances with respect to the contact and graphene resistances and small polarizations $P$ of the injector/detector electrodes.

In the limit of high impedance contacts ($R_C\rightarrow\infty$) Eq.~\ref{eq:Rnl2} reduces to Eq.~\ref{eq:Rnl} as expected and the spin signal decays exponentially with the distance. Reducing the value of the relaxation parameter (of the contact resistance for given $R_{sq}$ and $W$) the length dependence of the spin signal starts to deviate from the exponential form \cite{Kimura07}. In the case of finite $R_C$ and very long $\lambda$ ($R/\lambda\rightarrow 0$) and short devices ($L\ll \lambda$), the non-local resistance is $\pm(P^2 R_{sq}R^2)/(WL)$. Now, the spin signal is independent of $\lambda$, it depends on the injector detector spacing as $1/L$. Due to strong contact induced spin relaxation, spin transport is relevant only on the device length scale, the electrodes effectively cut the graphene flake with respect to the spin transport.

\begin{figure}[h!]
  \includegraphics[width=\columnwidth]{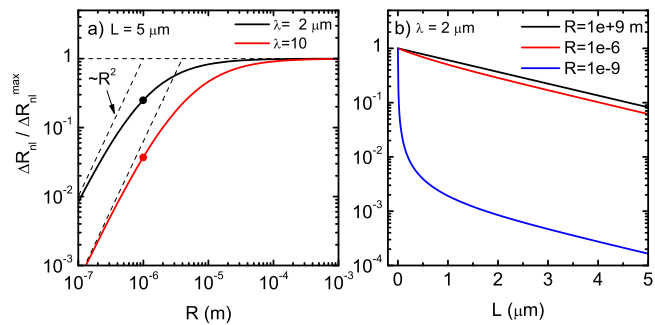}\\
  \caption{The variation of the spin valve signal (normalized to the maximum spin signal ideally available) as a function of: a) the relaxation parameter $R$ for two values of $\lambda=2,~10~\mu$m; b) injector-detector separation $L$ for $R=10^{+9},~10^{-6},~10^{-9}$~m.}
  \label{fig:RnlvsR}
\end{figure}

In Fig.~\ref{fig:RnlvsR}a) we plotted the variation of the non-local spin valve signal with respect to the relaxation parameter $R$ for a device with an injector-detector separation $L=5~\mu$m and two values for $\lambda$, 2 and 10~$\mu$m, respectively. The signal is normalized to the value corresponding to infinite contact resistance. For given graphene properties, a variable $R$ parameter represents a variable contact resistance $R_C$. The full circles in Fig.~\ref{fig:RnlvsR}a) correspond to $R=10^{-6}$~m (i.e., $R_C=1$~k$\Omega$ for typical values $R_{sq}=1$~k$\Omega$ and $W=1~\mu$m). Due to the conductivity mismatch, for $\lambda=2~\mu$m one measures only 25~\% of the totaly available signal in the ideal case of high impedance contacts. For $\lambda=10~\mu$m, only 3.7~\% of the ideal signal is measured. The strong contact induced spin relaxation regime is identified as the region where the signal is proportional to $R^2$. In Fig.~\ref{fig:RnlvsR}b) we show the variation of the spin signal with respect to the injector-detector separation for $\lambda=2\mu$m and three values of the relaxation parameter. For $R=10^{-9}$m and $L\leq\lambda$ the spin signal depends on the injector-detector separation as $1/L$.

\subsection{Graphene spin valve measurements}

\begin{figure}[h!]
  \includegraphics[width=\columnwidth]{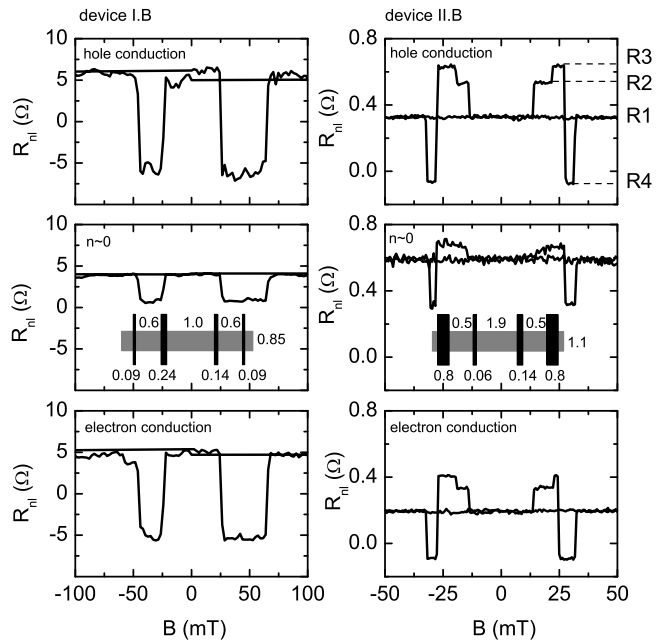}
  \caption{Spin valve measurements for a type~I device (left column) and a type~II device (right column), dimensions given in the insets in micrometers. For the Dirac point (middle graphs) the spin signals are smaller than for metallic regimes (densities of 2-3$\times$10$^{12}$cm$^{-2}$). The y-scales have the same span for each device. The resistance levels R1, R2, R3 and R4 observed for the device~II.B are due to the magnetic configuration of all four electrodes (see Fig.~\ref{fig:SVGT02Steps} and the text).}
  \label{fig:SVtypeIvsII}
\end{figure}

In this section we discuss the general characteristics of the spin valve measurements. In Fig.~\ref{fig:SVtypeIvsII} we show spin valve measurements for a type~I and a type~II device for three charge density regimes: zero charge density (Dirac neutrality point) and metallic hole and electron densities of 2-3$\times$10$^{12}$cm$^{-2}$.
Due to the different contact separations, carrier mobilities and metallic charge densities, a direct comparison between these measurements is not trivial. In general, we observe large signals (few ohms up to 50 ohms) for the type~I devices, whereas for the type~II the spin signals were a few tenths of an ohm up to a few ohms. We identify two reasons for this difference. First, for the type~I devices the contact resistances were in the order of 10 to 200~k$\Omega$, whereas for the type~II the contact resistances were always lower, of the order of 1 to 10~k$\Omega$. As already discussed the low impedance contacts induce significant spin relaxation, which severely reduces the magnitude of the spin signal. Second, the measurements are consistent with relatively high spin injection/detection efficiencies ($P$=10-20\%) in the case of high impedance contacts, whereas for low impedance contacts the efficiencies were of the order of 3-5\%.

\begin{figure}[h!]
  \includegraphics[width=.95\columnwidth]{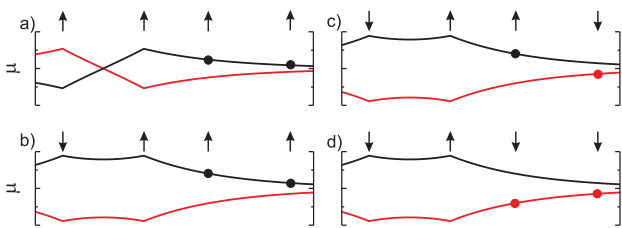}\\
  \caption{Schematic representation of the electrochemical potentials for the spin up and spin down channels as a function distance for the device II.B in Fig.~\ref{fig:SVtypeIvsII}. The magnetic configurations of all four ferromagnetic electrodes (which give the multiple resistance levels R1, R2, R3 and R4) and the electrode positions are indicated by the arrows.}
  \label{fig:SVGT02Steps}
\end{figure}

For the type~II device in Fig.~\ref{fig:SVtypeIvsII} we observe four resistance steps which we associate with the magnetization switching of all four electrodes, graphically shown in Fig.~\ref{fig:SVGT02Steps}. Let us consider that for high positive magnetic field the magnetization of all electrodes points in the "up" direction and that F2 injects spin up carriers in graphene. At F1, where the current flows from graphene to F1, spin down injection takes place. This partially cancels the spin up accumulation created by F2. Both F3 and F4 electrodes probe the spin up channel and a positive non-local voltage (resistance) is measured since F3 is closer to the injectors than F4, Fig.~\ref{fig:SVGT02Steps}a). Ramping down the field towards -50~mT the electrodes switch their magnetization in the reverse order of their width: F1, F4, F3 and F2. When F1 changes its magnetization orientation, both F1 and F2 inject spin up carriers and the total spin accumulation increases causing a positive change in $R_{nl}$, Fig.~\ref{fig:SVGT02Steps}b). Next, the detector F4 switches, probing the spin down channel, and the non-local voltage difference increases again, Fig.~\ref{fig:SVGT02Steps}c). At around -25~mT, F3 switches its magnetization. Now, both injectors inject spin up carriers whereas the detectors probe the spin down ones which causes $R_{nl}$ to become negative, Fig.~\ref{fig:SVGT02Steps}d). The last switch is that of F2. This configuration (not shown) is equivalent to the starting one with all magnetizations pointing down now and the starting level of non-local resistance is recovered.

The four changes in the non-local resistance point out that the spin relaxation length in graphene is at least of the order of the F1-F3 or F4-F2 distances ($\sim2.9~\mu$m for this sample) since otherwise the resistance step due to F1 or F4 switching their magnetization would not be observed. Additionally, the spin accumulation can extend underneath all contacts, which means that the aluminum oxide barriers are not too opaque nor too transparent. In the next sections we extract quantitative information regarding the spin relaxation length $\lambda$.

\subsection{Length dependence of the spin signal}

Eq.~\ref{eq:Rnl2} can be used to extract $\lambda$ from the length dependence of the spin signal. Ideally, all one has to do is to make a series of devices which should strictly meet the following conditions. First, the F1 and F4 electrodes do not contribute to the measurement, they are either placed at distances much larger than $\lambda$ with respect to F2 and F3 or they are replaced with non-magnetic ones. Second, F2 and F3 should have identical spin injection/detection efficiencies and resistances from device to device. In addition, all the graphene flakes should have identical transport properties (same carrier mobility) and the same width (in case that enhanced spin relaxation takes place at the device edges). The samples should be processed and measured in identical conditions, since it is not actually known how the substrate could influence the spin relaxation. If all these conditions are met, then the only variable left is the spacing between the injector and the detector electrodes. This has to be varied systematically, so that the application of Eq.~\ref{eq:Rnl2} would allow one to extract $\lambda$ and $P$. Experimentally some of these conditions can be fairly met (especially if several devices can be fitted on the same flake), except controlling the polarization of the electrodes and the reproducibility of contact resistance which are probably the most important parameters. In spite of these problems we have been able to show some consistent behavior for samples where the contact resistances were more uniform \cite{Tombros08}, see also Fig.~\ref{fig:T1T2}a).

\begin{figure}[h!]
  \includegraphics[width=0.95\columnwidth]{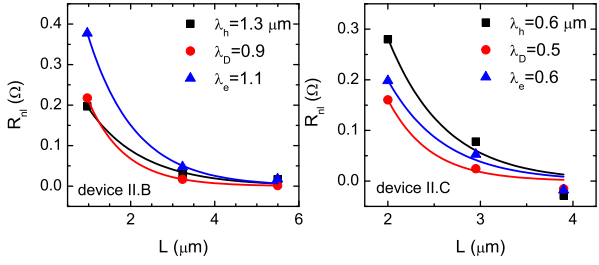}\\
  \caption{Length dependence of the spin signal as extracted from the magnitude of the multiple resistance levels (type II devices) for three conduction regimes: n$\sim0$ and hole/electron densities of 2-3$\times$10$^{12}$cm$^{-2}$. For these samples we calculated an average $R$ parameter of 1-2$\times$10$^{-6}$~m. The spin relaxation lengths are higher in the metallic regime (subscript $e$/$h$ for electron/hole conduction) than at the Dirac neutrality point (subscript $D$).}
  \label{fig:LDepSwitches}
\end{figure}

In this paragraph we show how the multiple resistance steps, which are due to ferromagnetic electrodes placed at various distances, can be used for a quantitative estimation of $\lambda$. For the all up magnetization configuration which gives the resistance level $R_1$ (Fig.~\ref{fig:SVtypeIvsII} and Fig.~\ref{fig:SVGT02Steps}a) the spin accumulation present at the position of the voltage probe F3 is just the sum of two spin accumulations of opposite sign produced by F2 and F1 which have decayed exponentially with the distance (F2-F3 and F1-F3). We neglect here the influence of the electrode F2 on the spin accumulation produced by F1 and detected by F3. The same arguments apply for the detector F4 but the exponential decays are for larger distances. Additionally, we note that some measurements (including this one) show a gate voltage dependent background resistance (which we denote as $R_0$). Note that, theoretically, the spin signals $R_2$ and $R_4$ should have identical values but opposite signs, see Fig.~\ref{fig:SVGT02Steps} b) and d). Therefore, we can write $R_1=[+R_{23}-R_{13}]-[+R_{24}-R_{14}]+R_0$, where $R_{23}$ is the resistance level produced by the F2-F3 injector-detector pair which are placed at the distance $L_{32}$ and so on. Applying a similar reasoning for $R_2$,$R_3$ and $R_4$ we obtain a system of four equations with five unknowns $R_{23},R_{13},R_{24},R_{14}$ and $R_0$. In this sample the F1-F3 and F2-F4 distances were virtually equal ($\sim2.9~\mu$m) so by taking $R_{13}=R_{24}$ we can solve the system of equations and extract the length dependence of the spin signal from a single measurement. The same analysis has been performed for another sample. The results are plotted in Fig.~\ref{fig:LDepSwitches} together with fits according to Eq.~\ref{eq:Rnl2}. The fits reveal a spin relaxation length of the order of $1~\mu$m, somewhat higher in the metallic regime than for the zero charge density regime.

\subsection{Modeling the contact resistance effects in the spin precession measurements}

\begin{figure}[h!]
  \includegraphics[width=0.9\columnwidth]{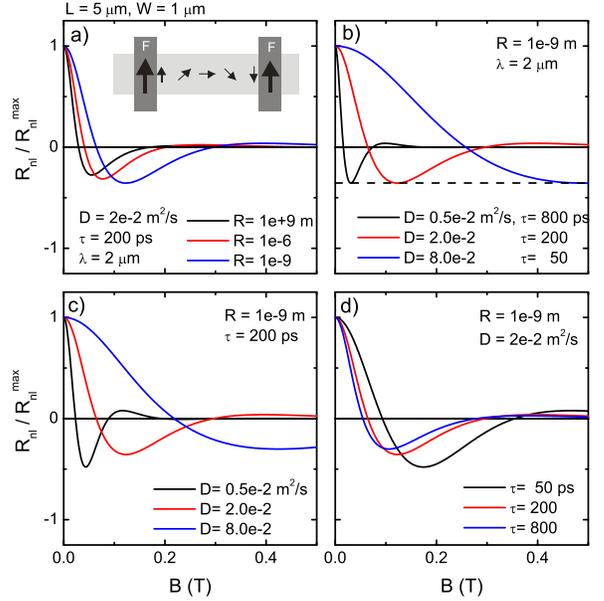}\\
  \caption{Calculations of the spin precession curves for various values of $R$, $D$ and $\lambda$. Only the parallel magnetization of the injector/detector electrodes case is shown for positive values of the perpendicular magnetic field.}
  \label{fig:ThPrec2}
\end{figure}

In this section we discuss the effect of the contact impedance on the spin precession measurements. Let us consider a spin valve device in which the ferromagnetic injector/detector electrodes have their respective magnetizations oriented parallel and a magnetic field $B_z$, smaller than what is required to pull the magnetization of ferromagnetic electrodes out of plane, is applied perpendicular to the sample plane. In this situation, the injected spins undergo precession (Hanle effect) while diffusing from the injector towards the detector, see the schematics of the inset in Fig~\ref{fig:ThPrec2}a). The detector signal oscillates with $B_z$.
The signal is maximum at $B_z=0$ (no precession), vanishes at fields which correspond to an average spin precession angle of 90 degrees and it is minimum when the spins have precessed an average angle of 180$^\circ$. In Fig.~\ref{fig:ThPrec2} we present several calculated precession signals for various values of the relaxation parameter $R$, the diffusion constant $D$ and the spin relaxation time $\tau$. All relevant parameters are indicated on the figures. To allow comparison the signals have been normalized to the maximum value. We refer to the Appendix for modeling details. Next, we analyze the shape of the curves with respect to their width, defined by the 90$^\circ$ precession angle, and the amount of overshoot, defined as the ratio between the minimum and the maximum values of the signal (180$^\circ$ and 0$^\circ$ precession). In Fig.~\ref{fig:ThPrec2}a) we plotted the precession signal for three values of $R$ and for fixed $D$ and $\tau$ (values close to the ones corresponding to our samples). With decreasing the relaxation parameter $R$, i.e. decreasing the contact resistance for given graphene properties, the precession curve becomes broader and the overshoot increases in magnitude. Of interest is to determine how $R,D,\tau$ influences the shape of the precession curve, and whether one can accurately extract the spin relaxation time in the case of very low impedance contacts, for example $R_C=1\Omega$, which for $R_{sq}=1$k$\Omega$ and $W=1\mu$m corresponds to $R=10^{-9}$m.

In Fig.~\ref{fig:ThPrec2}b) we show three curves for which $R=10^{-9}$m and the values for $D$ and $\tau$ are such that they give a constant spin relaxation length of 2~$\mu$m. This plot indicates that the overshoot depends only on $\lambda$. Extensive analysis reveals that for a given injector-detector distance $L$ and for $\lambda\lesssim L$, the overshoot depends only on $\lambda$ and $R$, it does not depend on the individual values of $D$ and $\tau$ as long as they produce the same $\lambda$. Therefore, for a given sample (for which $L$ and $R$ are known) the spin relaxation length can be estimated just by quantifying the amount of the overshoot. In Figs.~\ref{fig:ThPrec2}c) and d) we plotted spin precession signals by varying either $D$ or $\tau$ while keeping the other parameters constant. For a given $\tau$, the precession curve is wider for a higher $D$. This can be simply understood by the fact that for a higher $D$ (i.e. a higher carrier mobility), a stronger magnetic field is required to induce 90$^\circ$ precession since it takes less time for the spins to diffuse the fixed injector/detector distance. Fig.~\ref{fig:ThPrec2}d) shows wider precession curves for smaller spin relaxation times. For high relaxation times, the shape of the curves is essentially the same. This can be understood by the fact that most of the injected spins do not relax before reaching the detector and the angle of precession is mainly determined by the time it takes them to arrive at the detector, i.e. the diffusion constant and the device length. For low spin relaxation times, the spins following a long diffusive path have a high chance to relax before reaching the detector and, therefore, do not contribute to the measured signal. In this case, the spin signal is mainly determined by the spins following a short diffusive path, and which arrive faster at the detector (effectively having a higher diffusion constant). The overall behavior is that, for given device characteristics and dimensions, the precession measurements allow accurate extraction of the spin transport properties provided that the contacts do not induce strong spin relaxation, i.e. the relaxation parameter $R$ is comparable or higher than the spin relaxation length.

\subsection{Spin precession experiments}

\begin{figure}[h!]
  \includegraphics[width=0.8\columnwidth]{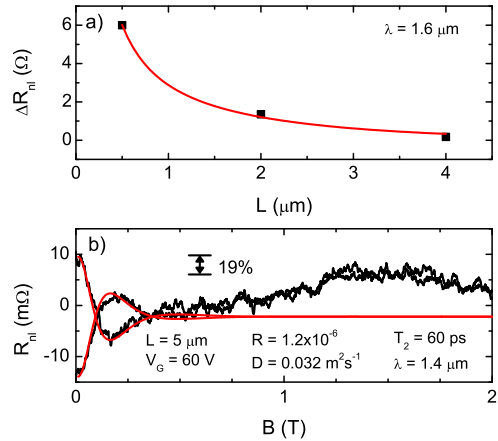}\\
  \caption{Measurements at 4.2~K in the metallic hole regime, type~I devices. a) length dependence of the spin valve signal and b) spin precession.}
  \label{fig:T1T2}
\end{figure}

In Fig.~\ref{fig:T1T2}a) we show the length dependence of the spin valve signal for three type~I devices made on the same graphene flake measured at 4.2~K in the metallic hole conduction regime $n(h)\sim2\times 10^{12}$~cm$^{-2}$. For this sample the contact resistances were more uniform (1-2~k$\Omega$). The measurements indicate a spin relaxation length of 1.6~$\mu$m.
In Fig.~\ref{fig:T1T2}b) we present a spin precession measurement for a 5~$\mu$m long device from the same batch of samples. The fit of the low $B$ field part of the precession curve indicates a spin relaxation length of 1.4~$\mu$m, in very close agreement with the length dependence measurement. This indicates that the longitudinal and transversal relaxation times (T1 and T2) are similar for these devices \cite{Tombros08}. Therefore, we argue that the precession measurements can be used to determine the spin transport properties ($\tau$,$\lambda$) of graphene. The advantage over the difficult to realize length dependence of the spin signal (due to the irreproducibility of the contact resistances or the spin injection/detection efficiencies) is obvious: the fabrication and proper characterization of a single device.

With increasing the perpendicular magnetic field above 0.5~T the magnetization of the Co electrodes is slowly pulled out of plane. For fields above 1.25~T the relative orientation of the magnetization of the Co electrodes is parallel (both magnetizations are completely out of plane). Now, the injected/detected spins are perpendicular to the two dimensional plane of the graphene layer and parallel to the external field (no precession takes place). We observe that the spin signal is somewhat lower than the zero field signal, when the orientation of the injected/detected spins is parallel with the sample plane. The difference in the magnitude of the signal corresponds to a relaxation time $\tau_\perp$ smaller by $\simeq 20$~\% than $\tau_\parallel$. Similar room temperature experiments not shown here support these low temperature spin anisotropy measurements. This indicates that the effective magnetic fields responsible for the spin relaxation lie mostly in the two dimensional plane of graphene. For a full discussion we refer to \cite{Tombros08}. However, we note that in the case of Rashba or Dresselhaus type spin orbit interaction the effective spin orbit fields lie exclusively in the graphene plane and, theoretical calculations indicate that $\tau_\perp=(1/2)\tau_\parallel$ \cite{Fabian07}.

\begin{figure}[h!]
  \includegraphics[width=\columnwidth]{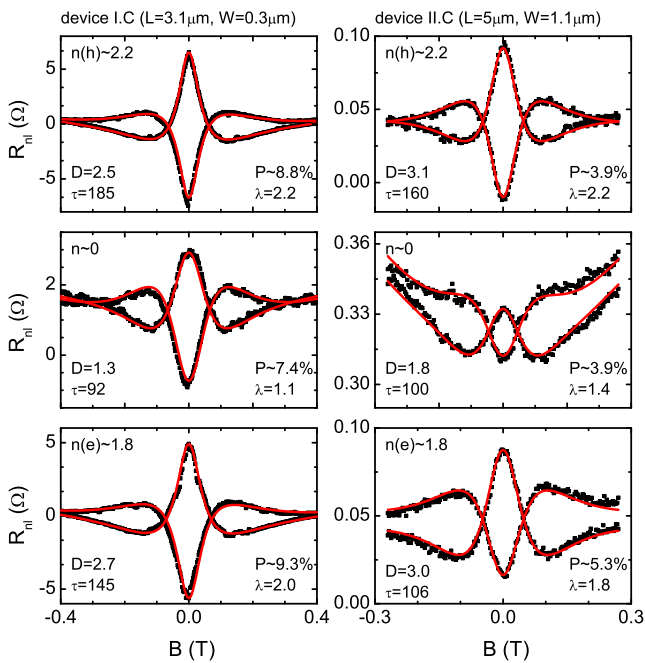}\\
  \caption{Room temperature spin precession measurements for a type~I device (left column) with $R_C$=20-40~k$\Omega$, $\mu=2.4\times 10^3$~cm$^2$/Vs and for a type~II device (right column) with $R_C$=1-2~k$\Omega$, $\mu=5\times 10^3$~cm$^2$/Vs in different conduction regimes. The carrier concentrations are expressed in units of $10^{12}$~cm$^{-2}$. The flake dimensions are indicated on the figure. The solid red lines through the data points are fits with the parameters indicated on the figures. $D$, $\tau$ and $\lambda$ are expressed in units of $10^{-2}$m$^2$/s, ps and $\mu$m respectively.}
  \label{fig:PrecIvsII}
\end{figure}

In Fig.~\ref{fig:PrecIvsII} we show spin precession measurements for a type~I (contact resistances of 20-40~k$\Omega$) and a type~II sample (contact resistances of 1-2~k$\Omega$) for similar charge carrier type and density. Even though the mobilities and widths of the two devices were different, the spin relaxation lengths are rather similar. This indicates that the presence of the aluminium oxide layer on top of graphene (type~I sample) does not influence the spin transport properties significantly.

Overall the values of the diffusion constant agree well with the ones extracted from the conductivity measurements (see \cite{Tombros07,Tombros08} for details), whereas the spin relaxation times are of the order of 100-200~ps. Only a few devices show relaxation times smaller than 100~ps. In all the measurements we observe that the spin transport properties depend on the gate voltage, i.e. on the charge density. Generally, for the same device we observe that both the spin diffusion constant $D$ and the spin relaxation time $\tau$ at the Dirac point are approximately a factor of two smaller than at high carrier densities, see Fig.~\ref{fig:PrecIvsII}. Since $D=v_F \tau_e/2$ (with $v_F$ the Fermi velocity and $\tau_e$ the momentum relaxation time), it seems that $\tau\propto\tau_e$. This behavior is expected for the Elliott-Yafet type spin scattering mechanism.

\subsection{Spin precession in graphene ribbons}

\begin{figure}[t!]
  \includegraphics[width=\columnwidth]{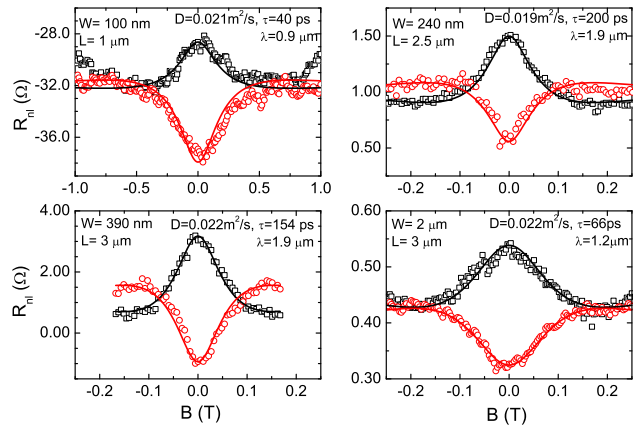}\\
  \caption{Room temperature spin precession measurements in the metallic hole conduction regime $n(h)\sim2\times 10^{12}$~cm$^{-2}$ for four type~III devices of different widths.}
  \label{fig:PrecTypeIII}
\end{figure}
In this section we discuss room temperature spin precession measurements for structures down to 100~nm in width (type~III devices). These measurements were performed in order to establish whether the short spin relaxation times may be due to enhanced spin flip processes taking place at the edges of the graphene flake. The fabrication procedure involved an oxygen plasma etching step and it was described in the Experimental section. The values of the contact resistances for this type of etched structures show the poor reproducibility compared with the type~I and II devices. We attribute this to the unknown residues produced during the etching process which hinders making proper contacts to graphene. In Fig.~\ref{fig:PrecTypeIII} we show a set of four precession measurements performed on different devices in the metallic hole conduction regime. In spite of similar charge transport properties, with carrier mobilities of the order to 2-2.5$\times 10^{3}$~cm$2$/Vs, the spread in the spin relaxation times is rather large. Generally, the spin valve signals at the Dirac point were too small to produce a useful precession curve. For some precession curves there are a few sets of parameters that describe reasonably the experiment. In that case we fixed the spin diffusion constant to the charge diffusion constant, leaving the spin relaxation time as the only relevant fitting parameter. In the order of increasing the graphene flake width (0.10, 0.24, 0.39 and 2~$\mu$m) the spin relaxation times read 40, 200, 154 and 60~ps respectively. No clear scaling behavior is found. Though the spin relaxation time for the narrowest device (100~nm) shows the lowest value of 40~ps, a definitive conclusion with respect to a presumably strong spin relaxation at the device edges cannot be drawn since spin relaxation times of 50-60~ps were measured for relatively wide flakes as well.

\section{Conclusions}

We have successfully realized all electrical injection and detection of spin accumulation in graphene at room temperature and 4.2~K. The conductivity mismatch problem has been partially overcome by the introduction of a thin aluminium oxide layer between the metallic Co electrodes and the semiconducting graphene. By proper modeling, we have shown that the spin relaxation via the contacts does not dominate the spin transport properties of our devices. From the length dependence of the spin valve signal and spin precession measurements, we have shown that the longitudinal and transversal spin relaxation times are similar. For the 2-5$\times 10^3$~cm$^2$/Vs carrier mobilities of our samples, we found spin relaxation times of 50-200~ps. Good agreement has been found between the spin diffusion constant and charge diffusion constant. The linear dependence of the spin relaxation time on the momentum scattering time indicates that the spin relaxation mechanism is Elliot-Yafet type. For carrier mobilities a factor of 100 higher than our samples, a spin relaxation length of the order of 100 micrometers is expected. The anisotropy in spin relaxation times reveal that the effective magnetic fields due to the spin orbit interaction lie mostly in the two dimensional graphene plane. In the range of carrier mobilities of our samples, the scattering at the devices edges, for structures down to 100~nm in width, and the presence of the aluminum oxide covering the graphene flakes in some experiments do not influence the relaxation times significantly.

\section*{Acknowledgments}

We would like to acknowledge B. Wolfs, S. Bakker, J. Baas and J. Buurma for technical support. This work was financed by MSC$^{plus}$, NanoNed, NWO (via a 'PIONIER' grant and a 'Top Talent' grant) and FOM.

\section*{Appendix. Modeling of spin relaxation induced by the contacts}

\begin{figure}[ht!]
  \includegraphics[width=.7\columnwidth]{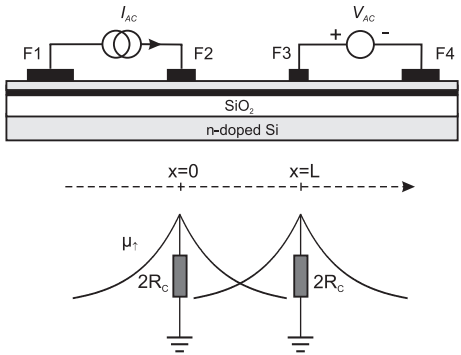}\\
  \caption{Modeling of the spin relaxation via the contacts. F1 and F4 are considered as non-magnetic.}
  \label{fig:SVR}
\end{figure}

We consider the geometry shown in Fig.~\ref{fig:SVR} with F1 and F4 being non-magnetic. The injector F2 and the detector F3 are placed at the distance $L$. In the frame of the two channel current model the current is carried by two independent channels, a spin up and a spin down one. In the absence of external magnetic fields the spin accumulation is constant in the $y$ and $z$ directions, only the $x$ component varies with the distance. To simplify the mathematical description we consider symmetric splitting of the electrochemical potentials of the two spin species and discuss only the spin up channel.
Neglecting the linear term due to charge current flow for $x<0$, the solutions for the spin up electrochemical potential in the $x$ direction are of the form
\begin{subequations}\label{eq:sol1}
\begin{eqnarray*}
\mu_\uparrow(x)&=&a\exp(+\frac{x}{\lambda})\label{eq:sol1.1},\quad\mbox{for $x<0$}\\
\mu_\uparrow(x)&=&b\exp(-\frac{x}{\lambda})+c\exp(+\frac{x}{\lambda}),\quad\mbox{for $0<x<L$} \label{eq:sol1.2}\\
\mu_\uparrow(x)&=&d\exp(-\frac{x}{\lambda}),\quad\mbox{for $x>L,$}\label{eq:sol1.3}
\end{eqnarray*}
\end{subequations}
Considering parallel orientation of the magnetization of the injector/detector electrodes and the same polarization $P$ for the injector/detector electrodes, we write the non-local resistance as
\begin{equation}\label{eq:Rnl1}
R_{nl}(L)=\frac{P}{I_{ac}}\frac{\mu_\uparrow(L)}{e}= \frac{P}{I_{ac}}\frac{d\exp{(-L/\lambda)}}{e}
\end{equation}
where we took into account the symmetric splitting $\mu_\uparrow(L)=-\mu_\downarrow(L)$ when applying Eq.~2 from \cite{JedemaAPL}. The constants $a,b,c,d$ are determined from the boundary conditions. The continuity of the spin up electrochemical potential at $x=0,L$ gives the following two equations
\begin{subequations}\label{eq:mu1}
\begin{eqnarray}
a&=&b+c\\
b\exp(-\frac{L}{\lambda})+c\exp(+\frac{L}{\lambda})&=&d\exp(-\frac{L}{\lambda}).
\end{eqnarray}
\end{subequations}

The spin relaxation via the contacts is taken into account when writing the spin current conservation equations as explained in the following. The ferromagnet F2 injects a spin polarized current in graphene via tunnel barriers of resistance R$_C$ with an efficiency $P$. The total injected spin current is $I_s(0)=P I_{ac}$ and the total spin up current available is half, $I_\uparrow(0)=P I_{ac}/2=I_x(0)$. The spin up current $I_x(0)$ diffuses to the left and to the right in graphene or flows back into the contact. The spin up current flowing in graphene is written in the form $I^\uparrow_G=\pm (\sigma/2) W(1/e) d\mu_\uparrow/dx$. Here $e$ is the electron charge, $\sigma$ and $W$ are the conductivity and the width of the
graphene flake. The (unwanted) spin up current through the contact is written as $I^\uparrow_{C}=(1/R^\uparrow_C)\mu_\uparrow/e \approx(2eR_C)^{-1} \mu_\uparrow$. Here we have considered $R^\uparrow_{C} (=R^\downarrow_{C})\approx 2R_C$, with $R_C^\uparrow$ the effective resistance the up spins encounter when going back into contact. The spin current conservation equations at $x=0,L$ read
\begin{subequations}\label{eq:Is1}
\begin{eqnarray}
P\frac{I_{ac}}{2}&=&a\frac{\sigma W}{2e\lambda}+b\frac{\sigma W}{2e\lambda}+ c\frac{\sigma W}{2e\lambda}+\frac{a}{2e R_C}\\
0&=&b\frac{\sigma W}{2e\lambda}\exp(-\frac{L}{\lambda})+c\frac{\sigma W}{2e\lambda}\exp(+\frac{L}{\lambda})\nonumber\\
&&-d\frac{\sigma W}{2e\lambda}\exp(-\frac{L}{\lambda})+\frac{d}{2e
R_C}\exp(-\frac{L}{\lambda})
\end{eqnarray}
\end{subequations}
Here we have used the fact that the spin current flowing away from the contact (the exponential decaying electrochemical potential from the contact point of view) is positive. The boundary conditions give a system of four equations with four unknowns from which the constants $a,b,c,d$ are extracted. Using $\sigma=1/R_{sq}$, we find the expression for $R_{nl}(L)$, given by Eq.~\ref{eq:Rnl2}.

In the presence of an external magnetic field applied in the $z$ direction the general solutions of the spin up electrochemical potential are of the form \cite{Tombros08t}
\begin{eqnarray}\label{eq:muZsol}
\left(\begin{array}{c} \mu_{\uparrow x} \\ \mu_{\uparrow y} \\ \mu_{\uparrow z} \end{array}\right)^{0,L}=
 &&A_\mp^{0,L}\left(\begin{array}{c} 0 \\  0 \\ 1 \end{array}\right)\exp(\pm K_1 x)+\\ \nonumber
 &&B_\mp^{0,L}\left(\begin{array}{c} 1 \\ -i \\ 0 \end{array}\right)\exp(\pm K_2^- x)+\\ \nonumber
 &&C_\mp^{0,L}\left(\begin{array}{c} 1 \\  i \\ 0 \end{array}\right)\exp(\pm K_2^+ x)
\end{eqnarray}
where the $-(+)$ signs correspond to the positive (negative) $x$ direction and $K_1$ and $K_2^\pm$ are given by $K_1=(\sqrt{D\tau})^{-1}=\lambda^{-1}$ and $K_2^{\pm}=\lambda^{-1}\sqrt{1 \pm i \omega \tau}$. Since the electrochemical potential cannot increase when $x$ approaches $-(+)\infty$, the solutions for $x<0$ ($x>L$) contain only the exponentially decaying terms, i.e. the terms containing $+K_{1,2}$ for $x<0$ and the terms $-K_{1,2}$ for $x>L$. For $0<x<L$ the solutions contain six terms so that the back flow of spins due to the presence of the detector is taken into account. There are in total twelve coefficients ($A_+^0$,$B_+^0$,$C_+^0$, $A_-^0$,$B_-^0$,$C_-^0$,$A_+^L$,$B_+^L$,$C_+^L$,$A_-^L$,$B_-^L$,$C_-^L$) that describe the solutions. These are found from the boundary conditions: the continuity of electrochemical potentials and the spin current conservation, laws which we apply for each $x,y$ and $z$ component. The continuity of the electrochemical potentials give in total six equations, three for injection and three for detection. The spin current conservation laws give another six equations and they are written in a similar manner as it was described above. Except the $x$ component $I_x(0)$ all the other spin current components $I_y(0),I_z(0),I_x(L),I_y(L),I_z(L)$ are null. The spin up electrochemical potential present at the detector ($x$ component) is given by
\begin{equation}
\mu_\uparrow(L)=B_-^L\exp{(-K_2^-L)}+C_-^L\exp{(-K_2^+L)}.
\end{equation}

The boundary conditions give a system of twelve equations with twelve unknowns. For given $\tau$ and $D$ we solve the system of equations numerically to find the coefficients $B_-^L$ and $C_-^L$ for each value of the magnetic field. $P$ is just a multiplication factor. We calculate a series of precession curves for a mesh given by different values of $\tau$ and $D$. From this set of curves we choose the one that shows the smallest deviation from the experiment by the least mean squares method. The parameters $\tau,D,P$ corresponding to this curve represent the best fit.


\begin{thebibliography}{99}

\bibitem{Naber07} W. J. M. Naber, S. Faez, and W. G. van der Wiel, J. Phys. D: Appl. Phys. \textbf{40}, R205 (2007)

\bibitem{Hill06} E. W. Hill, A. K. Geim, K. Novoselov, F. Schedin, and P. Blake, IEEE Trans. Magn. \textbf{42 (10)}, 2694 (2006)
\bibitem{Tombros07} N. Tombros, C. Jozsa, M. Popinciuc, H. T. Jonkman, and B. J. van Wees, Nature \textbf{448}, 571 (2007).
\bibitem{Cho07} S. Cho, Y.-F. Chen, and M. S. Fuhrer, Appl. Phys. Lett. \textbf{91}, 123105 (2007)
\bibitem{Nishioka07} M. Nishioka, and A. M. Goldman, Appl. Phys. Lett. \textbf{90}, 252505 (2007)
\bibitem{Ohishi07} M. Ohishi, M. Shiraishi, R. Nouchi, T. Nozaki, T. Shinjo, and Y. Suzuki, Jpn. J. Appl. Phys., Vol. \textbf{46}, No. 25, L605-L607 (2007)
\bibitem{Wang08} W. H. Wang, K. Pi, Y. Li, Y. F. Chiang, P. Wei, J. Shi, and R. K. Kawakami, Phys. Rev. B, Vol. \textbf{77}, 020402(R) (2008)
\bibitem{Tombros08} N. Tombros, S. Tanabe, A. Veligura, C. Jozsa, M. Popinciuc, H. T. Jonkman, and B. J. van Wees, Phys. Rev. Lett., \textbf{101}, 046601 (2008)
\bibitem{Jozsa08} C. Jozsa, M. Popinciuc, N. Tombros, H. T. Jonkman, and B. J. van Wees, Phys. Rev. Lett., \textbf{100}, 236603 (2008)
\bibitem{Goto08} H. Goto, A. Kanda, T. Sato, S. Tanaka, Y. Ootuka, S. Odaka, H. Miyazaki, K. Tsukagoshi, and Y. Aoyagi, Appl. Phys. Lett., \textbf{92}, 212110 (2008)
\bibitem{Jozsa09} C. Jozsa, M. Popinciuc, N. Tombros, H. T. Jonkman, and B. J. van Wees, Phys. Rev. B, \textbf{79}, 081402(R) (2009).
\bibitem{Shiraishi08} M. Shiraishi, M. Ohishi, R. Nouchi, T. Nozaki, T. Shinjo, and Y. Suzuki, arXiv:0810.4592
\bibitem{Han09a} W. Han, W. H. Wang, K. Pi, K. M. McCreary, W. Bao, Y. Li, F. Miao, C. N. Lau, and R. K. Kawakami, Phys. Rev. Lett. \textbf{102}, 137205 (2009)
\bibitem{Han09b} W. Han, K. Pi, W. Bao, K. M. McCreary, Y. Li, W. H. Wang, C. N. Lau, and R. K. Kawakami, Appl. Phys. Lett. \textbf{94}, 222109 (2009)


\bibitem{Novoselov04} K. S. Novoselov, A. K. Geim, S. V. Morozov, D. Jiang, Y. Zhang, S. V. Dubonos, I. V. Grigorieva, and A. A. Firsov, Science \textbf{306}, 666 (2004).
\bibitem{RiseofG} A. K. Geim, and K. S. Novoselov, Nature Mater. \textbf{6}, 183-191 (2007).
\bibitem{Bolotin08} K. I. Bolotin, K. J. Sikes, Z. Jiang, M. Klima, G. Fudenberg, J. Hone, P. Kim, and H. L. Stormer, Solid State Commun. \textbf{146}, 351 (2008)

\bibitem{Kane05} C. L Kane, and E. J. Mele, Phys. Rev. Lett. \textbf{95}, 226801 (2005)
\bibitem{Hernando06} D. Huertas-Hernando, F. Guinea, and A. Brataas, Phys. Rev. B \textbf{74}, 155426 (2006)
\bibitem{Min06} H. Min, J. E. Hill, N. A. Sinitsyn, B. R. Sahu, L. Kleinman, and A. H. MacDonald, Phys. Rev. B \textbf{74}, 165310 (2006)
\bibitem{Yao07} Y. Yao, F. Ye, X-L. Qi, S-C. Zhang, and Z. Fang, Phys. Rev. B \textbf{75}, 041401(R) (2007)
\bibitem{Hernando08} D. Huertas-Hernando, F. Guinea, and A. Brataas, arXiv:0812.1921v2 (2008)
\bibitem{Gmitra09} M. Gmitra, S. Konschuh, C. Ertler, C. Ambrosch-Draxl, and J. Fabian, arXiv:0904.3315v2 (2009)
\bibitem{Ertler09} C. Ertler, S. Konschuh, M. Gmitra, and J. Fabian, arXiv:0905.0424v2 (2009)
\bibitem{CastroNeto09} A. H. Castro Neto, and F. Guinea, Phys. Rev. Lett. 103, 026804 (2009)

\bibitem{Schmidt00} G. Schmidt, D. Ferrand, L. W. Molenkamp, A. T. Filip, and B. J. van Wees, Phys. Rev. B \textbf{62}, R4790 (2000)
\bibitem{Rashba00} E. I. Rashba, Phys. Rev. B \textbf{62}, R16267 (2000)
\bibitem{Fert01} A. Fert, and H. Jaffres, Phys. Rev. B \textbf{64}, 184420 (2001)
\bibitem{Smith01} D. L. Smith, and R. N. Silver, Phys. Rev. B, \textbf{64}, 045323 (2001)
\bibitem{Takahashi03} S. Takahashi, and S. Maekawa, Phys. Rev. B, \textbf{67}, 052409 (2003)
\bibitem{Staa08} A. van Staa, J. Wulfhorst, A. Vogel, U. Merkt, and G. Meier, Phys. Rev. B \textbf{77}, 214416 (2008)
\bibitem{Kimura07} T. Kimura, and Y. Otani, Phys. Rev. Lett. \textbf{99}, 196604 (2007)
\bibitem{Johnson88} M. Johnson, and R. H. Silsbee, Phys. Rev. B, \textbf{37}, 5312 (1988)
\bibitem{Jedema02} F. J. Jedema, H. B. Heersche, A. T. Filip, J. J. A. Baselmans, and B. J. van Wees, Nature \textbf{416}, 713 (2002)
\bibitem{vanSon87} P. C. van Son, H. van Kempen, and P. Wyder, Phys. Rev. Lett. \textbf{58} 2271 (1987)
\bibitem{JedemaAPL} F. J. Jedema, M. V. Costache, H. B. Heersche, J. J. A. Baselmans, and B. J. van Wees, Appl. Phys. Lett. \textbf{81}, 5162 (2002)
\bibitem{Tombros08t} N. Tombros, Ph.D. thesis, University of Groningen, The Netherlands
\bibitem{Fabian07} J. Fabian, A. Matos-Abiague, C. Ertler, P. Stano, and I. \v{Z}uti\'{c}, Acta Physica Slovaca \textbf{57}, 565-907 (2007).

\end{thebibliography}
\end{document}